\documentclass[12pt]{iopart}

\usepackage[squaren]{SIunits}
\usepackage{stmaryrd}
\usepackage{amsthm}
\usepackage{amssymb}
\usepackage{mathrsfs}

\usepackage{multirow}
\usepackage{color}
\usepackage{setspace}

\usepackage{graphicx}
\usepackage{subfig}

\usepackage{url}

\usepackage{hyperref}

\begin{document}

\title[Note on equatorial geodesics in circular spacetimes]{Note on equatorial geodesics in circular spacetimes}
\author{K.~Van Aelst$^{1}$}
\vspace{10pt}
\begin{indented}
\item[$^{1}$] Laboratoire Univers et Th\'eories, Observatoire de Paris, Universit\'e PSL, CNRS, Universit\'e de Paris, 92190 Meudon, France
\end{indented}
\ead{karim.van-aelst@obspm.fr}

\begin{abstract}
General results
on equatorial geodesics are exposed in the case of
circular spacetimes featuring an equatorial reflection symmetry.
The way the
geodesic equation equivalently
rewrites in terms of an effective potential is explicitly recalled for circular and non-circular equatorial geodesics.
This provides a practical tool
to discuss their stability.
Statements are illustrated in
Kerr spacetime.
\end{abstract}


\section{Introduction}

One of the first reference works on equatorial geodesics date back to the investigations of Boyer and Price~\cite{Boyer_Price_geods}, who focused on the Kerr metric just before the seminal articles
of Carter on general Kerr geodesics
~\cite{Carter_Analytic_extension,Carter_Global_structure_Kerr,Carter_Hamilton_Jacobi}.
Later on, Bardeen~\cite{Bardeen_stability}, Press and Teukolsky~\cite{Bardeen_rot_bh} considered equatorial geodesics in circular spacetimes.
A defining property of the latter is to admit a quasi-isotropic coordinate system, i.e. a coordinate system~$(t,r,\theta,\phi)$ with respect to which the line element writes
\begin{eqnarray}
\label{eq_circular_spacetime}
ds^{2} = - N^2 dt^2 + A^2 \left( dr^2 + r^2 d\theta^2 \right) + B^2 r^2 \sin^2 \theta \left( d\varphi - \omega dt \right)^2,
\end{eqnarray}
where the metric functions~$N$,~$A$,~$B$ and~$\omega$ only depend on coordinates~$r$ and~$\theta$.

At the time, these authors
were interested in the precession of the periapsis of bounded orbits, stability of thin dust disks and other processes around rotating black holes.
In regard of the highly accurate observations realized by instruments like GRAVITY~\cite{GRAVITY,GRAVITY_redshift_S2,GRAVITY_motion_ISCO} and the Event Horizon Telescope~\cite{EHT,EHT_Shape_SgrA,EHT_Shadow_M87}, such investigations are still essential today, e.g. to examine orbits of stars, accretion disks around
compact objects and the images they produce.
Contemporary discussions on these topics can
be found
in references~\cite{Eric_intro_relat_stars, equat_circu_Kerr, circular_rotating_boson_star, Gralla_shadow_rings, geods_shadow_GB}.

Circular spacetimes~(\ref{eq_circular_spacetime})
represent a large subclass of stationary and axisymmetric spacetimes (the corresponding Killing vectors being~$\partial_{t}$ and~$\partial_{\phi}$) possessing the additional symmetry~$(t,\phi)~\mapsto~(-t,-\phi)$.
The present note further assumes the equatorial reflection symmetry
\begin{eqnarray}
\label{eq_equatorial_symmetry}
\forall \mu,\nu,\ g_{\mu\nu}(r,\pi/2 - \theta) = g_{\mu\nu}(r,\theta).
\end{eqnarray}

Such metrics notably include the Kerr family%%%
\footnote{%%%
The corresponding expressions of the metric functions~$N$,~$A$,~$B$ and~$\omega$ (and the transformation from the usual Boyer-Lindquist coordinates to quasi-isotropic coordinates) are given in appendix~\ref{appdx_Kerr_QI}; reference~\cite{quasi_isotropic_Kerr} may also be consulted.
}%%%
,
numerical black hole metrics~\cite{Einstein_SU2YM_2, rotating_EdGB_2, Delgado_spin_Horn_Bonnet}, models of rotating neutron stars~\cite{circular_rotating_proto_neutron_stars} and boson stars~\cite{circular_rotating_boson_star}.
Yet, note for instance that Kerr-Newman-Taub-NUT metric~\cite{Kerr_Taub_NUT_Shadow, Kerr_Taub_NUT_charged_test_particles, Kerr_Taub_NUT_circular_orbits}
does not fall into this family as
equatorial symmetry~(\ref{eq_equatorial_symmetry}) cannot hold in presence of the gravitomagnetic parameter.
One may consult~\cite{Carter_killing_ortho_transitive, Carter_Houches, Heusler_uniqueness_book, Eric_intro_relat_stars, Eric_Silvano_noncircular} and references therein for further technical details on circular spacetimes and comments on their physical relevance.

\section{Conservation equations}
\label{section_conservation}

This section recalls the elements relevant in investigating the existence and properties of the trajectory of a free massive (resp. massless) particle in the equatorial plane of a circular spacetime.
To effectively search for such a trajectory, its parametrization is set to be the only one whose corresponding tangent vector is the 4-momentum~$p$ of the particle.
More explicitly, one looks for a timelike (resp. null), future-oriented curve
\begin{eqnarray}
\label{eq_curve}
\mathcal{C}:\ \lambda \mapsto \left( x^{\mu}(\lambda) \right) = \left( t(\lambda),\ r(\lambda),\ \theta(\lambda),\ \phi(\lambda) \right)
\end{eqnarray}
such that the 4-momentum of the particle is~$p^{\mu} = \dot{x}^{\mu}$, where a dot denotes differentiation with respect to the parameter~$\lambda$.
In addition, the particle is free if and only if its 4-momentum is parallely transported along its trajectory~$\mathcal{C}$:
\begin{eqnarray}
\label{eq_geodesic}
\nabla_{p}p = 0,
\end{eqnarray}
so that~$\mathcal{C}$ is an affinely parametrized geodesic by definition.

The geodesic equation~(\ref{eq_geodesic}) implies that the mass
\begin{eqnarray}
\label{eq_mass_conservation}
m = \sqrt{- p^{2}}\ \ \text{ is conserved along } \mathcal{C}.
\end{eqnarray}

In particular, if~$m > 0$,~$\lambda$ is necessarily the curvilinear abscissa (i.e. proper time)~$\tau$ along~$\mathcal{C}$ divided by~$m$.

In a stationary and axisymmetric spacetime such as~(\ref{eq_circular_spacetime}), equation~(\ref{eq_geodesic}) also implies conservation of the Killing energy and angular momentum:
\begin{eqnarray}
\label{eq_E_conservation}
E = - \partial_{t} \cdot p\ \ \text{ is conserved along } \mathcal{C}, \\
\label{eq_L_conservation}
L = \partial_{\phi} \cdot p\ \ \text{ is conserved along } \mathcal{C}.
\end{eqnarray}

Such quantities are actual observables only if the particle ever reaches spacelike infinity, where they are the energy and angular momentum effectively measured by a zero angular momentum observer (ZAMO)%%%
\footnote{%%%
The ZAMO are characterized by a 4-velocity colinear to~$\nabla t$; as a result, one may check that the ZAMO are not freely falling, yet they fulfill property~(\ref{eq_L_conservation}) with~$L = 0$, hence their name.%%%
}%%%
.
In this case,~$E \geq 0$ necessarily.

Finally, the trajectory is requested to be equatorial:
\begin{eqnarray}
\label{eq_equatorial_condition}
\theta = \pi/2\ \ \text{ is conserved along } \mathcal{C},
\end{eqnarray}
which implies~$p^{\theta} = 0$%%%
\footnote{%%%
This is equivalent to~$p_{\theta} \neq 0$ in quasi-isotropic coordinates~(\ref{eq_circular_spacetime}).%%%
}.

The conservation equations~(\ref{eq_mass_conservation}),~(\ref{eq_E_conservation}),~(\ref{eq_L_conservation}) and~(\ref{eq_equatorial_condition}) are thus four necessary conditions for a curve~$\mathcal{C}$ to describe an equatorial trajectory of a free particle.

\section{Non-circular geodesics}
\label{section_noncircular}

For non-circular orbits, i.e. for any trajectory such that~$p^{r} \neq 0$%%%
\footnote{%%%
This is equivalent to~$p_{r} \neq 0$ in quasi-isotropic coordinates~(\ref{eq_circular_spacetime}).
}
almost everywhere%%%
\footnote{%%%
Radial momentum $p^{r}$ may only cancel at periapsis and apoapsis, when they exist.%%%
},
these four conservation equations are sufficient:
they imply the geodesic equation~(\ref{eq_geodesic}) for the following reasons.
The Killing equation for~$\partial_{t}$
\begin{eqnarray}
\label{eq_Killing}
\left[ \nabla_{\mu} \partial_{t} \right]_{\nu} + \left[ \nabla_{\nu} \partial_{t} \right]_{\mu} = 0
\end{eqnarray}
and the Killing energy conservation~(\ref{eq_E_conservation}) establish the covariant~$t$ component of the geodesic equation:
\begin{eqnarray}
\label{eq_geod_t}
\left[ \nabla_{p}p \right]_{t} = \partial_{t} \cdot \nabla_{p}p = \nabla_{p} \left( \partial_{t} \cdot p \right) = - \nabla_{p}E = 0.
\end{eqnarray}

The analogous argument for~$\partial_{\phi}$ and~(\ref{eq_L_conservation}) yield~$\left[ \nabla_{p}p \right]_{\phi} = 0$.

Besides, the covariant~$\theta$ component of the geodesic equation also vanishes:
\begin{eqnarray}
\label{eq_geod_theta}
\left[ \nabla_{p}p \right]_{\theta} = A^{2} r^{2} \left[ \nabla_{p}p \right]^{\theta} =  A^{2} r^{2} p^{\mu} p^{\nu} \Gamma^{\theta}_{\phantom{\theta}\mu\nu} = 0
\end{eqnarray}
since~$\Gamma^{\theta}_{\phantom{\theta}r\theta}$ gets multiplied by~$p^{\theta}=0$, and all the other Christoffel symbols~$\Gamma^{\theta}_{\phantom{\theta}\mu\nu}$ vanish in the equatorial plane as sums of terms that are proportional either to~$\cos\theta$, or to some angular derivative of the metric~$\partial_{\theta}g_{\mu\nu}$, which is necessarily zero under the natural assumption of equatorial symmetry~(\ref{eq_equatorial_symmetry}).
When the latter does not hold, e.g. in Kerr-Newman-Taub-NUT spacetime, the covariant~$\theta$ component of the geodesic equation might not come out so simply: it may require alternative constraints on the metric components,
or invoke some of the other conservation equations.
It is indeed interesting to note that each of the above three covariant components of the geodesic equation
do not require any of the two other conservation equations to be derived.
As a result, any trajectory satisfying conservation of~$E$ (resp.~$L$, resp.~$\theta=\pi/2$ when equatorial symmetry holds) always satisfies the covariant~$t$ (resp.~$\phi$, resp.~$\theta$) component of the geodesic equation.

Finally, mass conservation~(\ref{eq_mass_conservation}) rewrites as
\begin{eqnarray}
\label{eq_nabla_mass}
0 = \nabla_{p}\left( p \cdot p \right) = p^{\mu} \left[ \nabla_{p}p \right]_{\mu} = p^{r} \left[ \nabla_{p}p \right]_{r},
\end{eqnarray}
which implies~$\left[ \nabla_{p}p \right]_{r} = 0$ since~$p^{r} \neq 0$ almost everywhere.

Before treating the circular case~$p^{r} = 0$, it is very useful to note that equations~(\ref{eq_E_conservation}),~(\ref{eq_L_conservation}) and~(\ref{eq_equatorial_condition}) allow to rewrite the mass conservation equation~(\ref{eq_mass_conservation}) as a familiar first order ordinary differential equation on the radial coordinate function~$r$:
\begin{eqnarray}
\label{eq_radial}
\frac{\dot{r}^{2}}{2} + \mathcal{V}(r,m,E,L) = 0,
\end{eqnarray}
where the effective potential~$\mathcal{V}$ is defined as
\begin{eqnarray}
\label{eq_potential}
\mathcal{V}(r,m,E,L) = \frac{1}{2A^{2}}\left[ m^{2} - \left( \frac{E - \omega L}{N} \right)^{2} + \left( \frac{L}{B r} \right)^{2} \right].
\end{eqnarray}

Since~$\lambda = \tau/m$ for massive particles, note that~(\ref{eq_radial}) rewrites as
\begin{eqnarray}
\label{eq_radial_per_mass}
\frac{1}{2} \left( \frac{dr}{d\tau} \right)^{2} + \mathcal{V}(r,1,\bar{E},\bar{L}) = 0,
\end{eqnarray}
where~$\bar{E} = E/m$ and~$\bar{L} = L/m$, so that the trajectories of free massive particles only depend on their Killing energy and angular momentum per unit mass.

Based on the explicit form~(\ref{eq_radial}) of the mass conservation equation, conditions on~$\mathcal{V}$ and its partial derivatives will also allow to characterize the circular geodesics (see section~\ref{section_circular}) and study their stability (see section~\ref{section_stability}).
So far, simply note that non-circular geodesics necessarily satisfy%%%
~$\mathcal{V} < 0$ almost everywhere, while circular geodesics necessarily satisfy%%%
~$\mathcal{V} = 0$ everywhere.

Finally, the procedure to explicitly construct all non-circular equatorial trajectories of free particles is to first pick an initial radial coordinate~$r_{0}$, a Killing energy~$E$, a Killing angular momentum~$L$ and a mass~$m$ such that~$\mathcal{V}(r_{0},m,E,L) < 0$ and~$E - \omega(r_{0}) L > 0$ to guarantee that the trajectory is initially causal future-oriented (see equation~(\ref{eq_t}) below); in particular, this necessarily requires~$E > 0$ if~$r_{0}$ is outside the ergoregion.
Recall that the latter is the domain over which the pseudo-stationary Killing vector~$\partial_{t}$ is non-timelike, i.e.~$N^{2} \leq (\omega B r \sin\theta)^{2}$.
Thus, realistic observers
can no longer have constant spatial coordinates.
Theoretically, this region allows to use particles to extract rotational energy from a black hole, as described by the Penrose process.
In practice, the latter is not
efficient enough to be significantly involved in
astrophysical processes such as the relativistic jets emerging e.g. from quasars, although this used to be conjectured.
Yet higher efficiencies might be reached around other objects, such as naked singularities or wormholes, or through more elaborate avatars of the process, such as the collisional Penrose process or superradiance.
One may consult~\cite{Abramowicz_review_accretion, Cardoso_review_Superradiance} and references therein for discussions of these topics.

Secondly,
the right-hand side of
\begin{eqnarray}
\label{eq_radial_sqrt}
\dot{r} = \pm \sqrt{-2 \mathcal{V}(r,m,E,L)}
\end{eqnarray}
is sufficiently regular for equation~(\ref{eq_radial_sqrt}) to admit a unique solution~$\lambda \mapsto r_{s}(\lambda)$ once the~$\pm$ sign is chosen (to determine whether the initial direction is ingoing or outgoing).
Conservation of~$E$ and~$L$ then provide the solutions for~$t$ and~$\phi$:
\begin{eqnarray}
\label{eq_t}
t_{s}(\lambda) = \int \frac{E - \omega(r_{s})L}{N(r_{s})^{2}} d\lambda, \\
\label{eq_phi}
\phi_{s}(\lambda) = \int \left[ \frac{L}{(B(r_{s}) r_{s})^{2}} + \omega(r_{s})\frac{E - \omega(r_{s})L}{N(r_{s})^{2}} \right] d\lambda,
\end{eqnarray}
where the second argument~$\theta_{s}=\pi/2$ of the metric functions is omitted without loss of clarity.
Note that frame dragging can be read off from relation~(\ref{eq_phi}): radial free fall from infinity (hence~$L=0$) does not remain radial in a rotating spacetime as~$\phi_{s}$ receives contribution from the non-vanishing metric function~$\omega$.

\section{Circular geodesics}
\label{section_circular}

As mentioned earlier,~$p^{r} = 0$ (circular orbit) forbids to establish the radial geodesic equation from mass conservation, which is then redundant as a linear combination of the~$E$ and~$L$ conservation equations (i.e. the covariant~$t$ and~$\phi$ geodesic equations).
Therefore, one additional equation is missing to realize a geodesic.
Indeed, there are so far multiple solutions to the problem~\{(\ref{eq_E_conservation}),~(\ref{eq_L_conservation}),~(\ref{eq_equatorial_condition}),~$p^{r} = 0$\} (which implies mass conservation): for any~$E$,~$L$ and~$r_{0}$, the curve
\begin{eqnarray}
\label{eq_circular_curve}
\lambda \mapsto \left( \frac{E - \omega_{0} L}{N_{0}^{2}} \lambda,\ r_{0},\ \pi/2,\  \left[ \frac{L}{(B_{0}r_{0})^{2}} +   \omega_{0} \frac{E - \omega_{0} L}{N_{0}^{2}} \right] \lambda \right)
\end{eqnarray}
is circular (an index~$0$ means that the metric function is evaluated at~$r_{0}$, in the equatorial plane) with conserved Killing energy~$E$ and angular momentum~$L$.
Additionally, one may simply require~$E \geq \omega_{0}L + N_{0} \vert L \vert/(B_{0}r_{0})$ (which is always positive outside the ergoregion) to describe a causal future-oriented trajectory; this amounts to requiring that mass conservation~$\mathcal{V}(r_{0},m,E,L) = 0$ holds for a real constant~$m$ (i.e.~$m^{2} > 0$).

All these circular orbits are distinct solutions to the same problem, but at most one of them is a geodesic, sometimes none (intuitively, the other orbits are accelerated inward if they rotate ``faster'' than a geodesic to compensate for the centrifugal effect, outward otherwise).
To search for a geodesic among them, one obtains an additional prescription from differentiating equation~(\ref{eq_radial}):
\begin{eqnarray}
\label{eq_diff_eq_radial}
\dot{r} \left[ \ddot{r} + \mathcal{V}'(r,m,E,L) \right] = 0,
\end{eqnarray}
where~$'$ denotes differentiation with respect to the first argument (the radial coordinate).

Of course, for any circular orbit~(\ref{eq_circular_curve}), equation~(\ref{eq_diff_eq_radial}) holds because~$\dot{r} = 0$, while it implies
\begin{eqnarray}
\label{eq_diff_eq_radial_noncircular}
\ddot{r} + \mathcal{V}'(r,m,E,L) = 0,
\end{eqnarray}
for all non-circular geodesics.
The missing condition to realize a circular geodesic at~$r_{0}$ is then obtained by requiring equation~(\ref{eq_diff_eq_radial_noncircular}) to hold even in the circular limit, i.e. when~$\ddot{r} = 0$, yielding
\begin{eqnarray}
\label{eq_circular_geod_condition}
\mathcal{V}'(r_{0},m,E,L) = 0.
\end{eqnarray}

\begin{figure}
    \begin{center}
        \subfloat[Effective potentials of timelike orbits with respect to~$\bar{r} = r/r_{\mathcal{H}}$ for different Killing energy and angular momentum per unit mass (only the ratios~$\bar{E} = E/m$ and~$\bar{L} = L/m$ are relevant in the massive case). The bottom red (resp. top blue) curve corresponds to the Killing energy and angular momentum of the timelike circular geodesic at~$4 r_{\mathcal{H}}$ (resp.~$15 r_{\mathcal{H}}$) while the other red (resp. blue) curve has same~$\bar{L}$ but a slightly smaller (resp. greater)~$\bar{E}$.]
        {
            \makebox[\textwidth][c]{
                \includegraphics[width=0.66\textwidth]{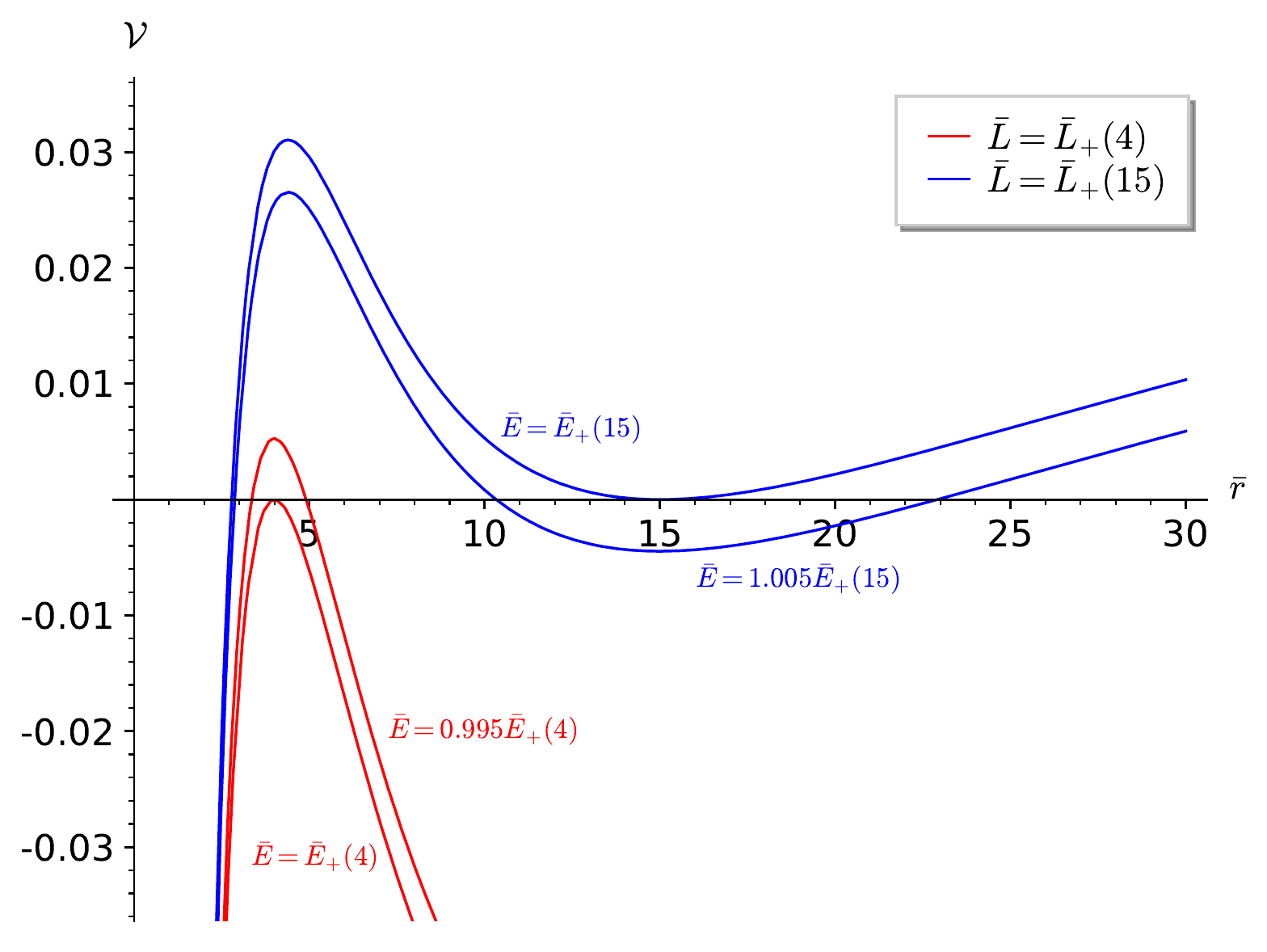}
                \label{fig_pot_Kerr}
            }
        } \\
        \subfloat[The red (resp. blue) dashed circle corresponds to the zero of the bottom red (resp. top blue) curve of figure~\ref{fig_pot_Kerr}. The red (resp. blue) solid line is ruled by the right, negative, unbounded (resp. bounded) branch of the top red (resp. bottom blue) curve of figure~\ref{fig_pot_Kerr}. Plot realized with the ray-tracing code~\emph{GYOTO}~\cite{Gyoto}.]
        {
            \makebox[\textwidth][c]{
                \includegraphics[clip, trim=2.5cm 0.5cm 3cm 1cm, width=0.60\textwidth]{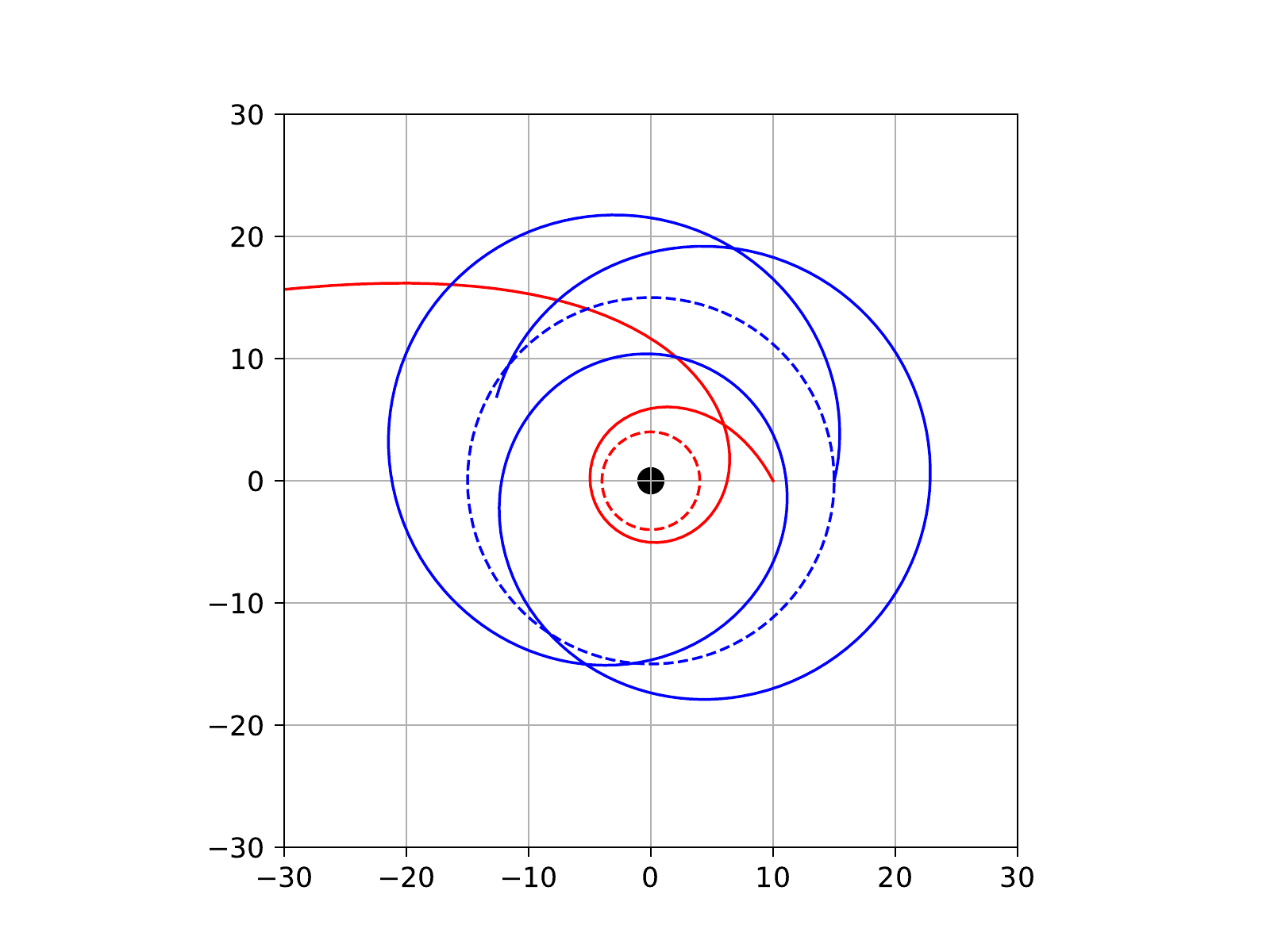}
                \label{fig_geods_Kerr}
            }
        }
    \end{center}
\caption{Potentials and spatial projections of timelike circular and non-circular equatorial geodesics in a Kerr spacetime~($a/M \simeq 0.52$).}
\label{fig_pot_geods_Kerr}
\end{figure}

On figure~\ref{fig_pot_Kerr} for instance, circular geodesics exist at the zeros of the bottom red and top blue curves since the latter cancel in a stationary way (the corresponding geodesics are respectively represented by the red and blue dashed lines in figure~\ref{fig_geods_Kerr}), while the zeros of the two other curves can only correspond to accelerated circular orbits or to the periapsis and apoapsis of a non-circular geodesic; more precisely, the smallest (resp. greatest) zero of the top red curve is the unique apoapsis (resp. periapsis) of a geodesic reaching the event horizon (resp. infinity, as represented by the red solid line in figure~\ref{fig_geods_Kerr}) while the smallest (resp. greatest) zero of the bottom blue curve is the periapsis (resp. apoapsis) of a bounded geodesic (represented by the blue solid line in figure~\ref{fig_geods_Kerr}).
Note here that, in the asymptotically flat case,~$N \rightarrow 1$,~$A \rightarrow 1$,~$B \rightarrow 1$ and~$\omega \rightarrow 0$ at infinity, so that~$\mathcal{V}(r,m,E,L) \rightarrow (1 - E^{2})/2$; therefore, a non-circular unbounded geodesic (such as the one represented by the red solid line on figure~\ref{fig_geods_Kerr} ruled by the right negative branch of the top red curve of figure~\ref{fig_pot_Kerr}) requires~$E > 1$.

To show that the additional condition~(\ref{eq_circular_geod_condition}) allows to establish the~$r$ geodesic equation, define
\begin{eqnarray}
\label{eq_W}
\mathcal{W}(r,m,E,L) = m^{2} - \left( \frac{E - \omega L}{N} \right)^{2} + \left( \frac{L}{B r} \right)^{2},
\end{eqnarray}
so that~$\mathcal{V}(r,m,E,L) = \mathcal{W}(r,m,E,L)/(2A^{2})$, and hence
\begin{eqnarray}
\label{eq_potential_prime}
\mathcal{V}'(r_{0},m,E,L) &= \left( \frac{1}{2A^{2}} \right)'_{0} \mathcal{W}(r_{0},m,E,L) + \frac{\mathcal{W}'(r_{0},m,E,L)}{2A^{2}_{0}} \\
&= \frac{\mathcal{W}'(r_{0},m,E,L)}{2A^{2}_{0}},
\end{eqnarray}
since mass conservation~$\mathcal{V}(r_{0},m,E,L) = 0$ is equivalent to~$\mathcal{W}(r_{0},m,E,L) = 0$.
Therefore, condition~(\ref{eq_circular_geod_condition}) is equivalent to
\begin{eqnarray}
\label{eq_circular_geod_condition_explicit}
0 &= \frac{\mathcal{W}'(r_{0},m,E,L)}{2} \\
&= \frac{E - \omega_{0} L}{N_{0}^{2}} \left[ (E - \omega_{0} L)\frac{N'_{0}}{N_{0}} + L \omega'_{0} \right] - \left( \frac{L}{B_{0} r_{0}} \right)^{2} \left( \frac{B'_{0}}{B_{0}} + \frac{1}{r_{0}} \right).
\end{eqnarray}

Then, injecting the explicit expressions
\begin{eqnarray}
\label{eq_L_circu_p}
L = B_{0}^{2} r_{0}^{2} \left( p^{\phi} - \omega_{0} p^{t} \right), \\
\label{eq_E_circu_p}
E = N_{0}^{2} p^{t} + \omega_{0} L,
\end{eqnarray}
into~(\ref{eq_circular_geod_condition_explicit}) immediately yields
\begin{eqnarray}
\label{eq_geod_r}
0 &= p^{t} \left[ N_{0} N'_{0} p^{t} + B_{0}^{2} r_{0}^{2} \left( p^{\phi}\!-\!\omega_{0} p^{t} \right) \omega'_{0} \right] - B_{0}^{2} r_{0}^{2} \left( p^{\phi}\!-\!\omega_{0} p^{t} \right)^{2} \left( \frac{B'_{0}}{B_{0}}\!+\!\frac{1}{r_{0}} \right)     \\
&= A_{0}^{2} \left[ \Gamma^{r}_{\phantom{r}tt} (p^{t})^{2} + 2\Gamma^{r}_{\phantom{r}t\phi} p^{t} p^{\phi} + \Gamma^{r}_{\phantom{r}\phi\phi} (p^{\phi})^{2} \right]      \\
&= A_{0}^{2} \Gamma^{r}_{\phantom{r}\mu\nu} p^{\mu} p^{\nu} = A_{0}^{2} \left[ \nabla_{p}p \right]^{r} = \left[ \nabla_{p}p \right]_{r}.
\end{eqnarray}

Therefore, circular geodesics are precisely the circular orbits~(\ref{eq_circular_curve}) that continue property~(\ref{eq_diff_eq_radial_noncircular}) of non-circular geodesics.
Gathered with the result of section~\ref{section_noncircular}, the circular and non-circular free trajectories are characterized as follows.
A curve~$\mathcal{C}:\ \lambda \mapsto \left( x^{\mu}(\lambda) \right)$ describes an equatorial free particle with 4-momentum~$p^{\mu} = \dot{x}^{\mu}$ if and only if it satisfies the three conservation equations~\{(\ref{eq_E_conservation}),~(\ref{eq_L_conservation}),~(\ref{eq_equatorial_condition})\} and either one of the two following conditions:
\begin{itemize}
\item the fourth conservation equation~(\ref{eq_mass_conservation}) and~$\mathcal{V} \neq 0$ almost everywhere on~$\mathcal{C}$ (in this case,~$\mathcal{C}$ is non-circular);
\item~$\mathcal{V} = 0$ and~$\mathcal{V}' = 0$ everywhere on~$\mathcal{C}$ (in this case,~$\mathcal{C}$ is circular).
\end{itemize}

Finally, to explicitly construct all free circular equatorial trajectories, recall that all the curves~(\ref{eq_circular_curve}) satisfying~$E \geq \omega L  + N \vert L \vert/(Br)$ (the indices~$0$ are now removed although all the statements in the remaining of the section will only apply to circular orbits) are very good candidates because they satisfy the three conservation equations~\{(\ref{eq_E_conservation}),~(\ref{eq_L_conservation}),~(\ref{eq_equatorial_condition})\} together with~$\mathcal{V}(r,m,E,L) = 0$, since this relation is the definition of~$m$ for a circular orbit at~$r$.
It thus only remains to derive which final constraint emerges from requiring~$\mathcal{V}'(r,m,E,L) = 0$ (in which~$m$ actually does not appear because of~$\mathcal{V}(r,m,E,L) = 0$).
This constraint will first be formulated in terms of the ``signed norm''~$V$ of the spatial velocity~$v$ measured by the ZAMO, defined below.
From~$V$ will then be deduced the values of~$E$ and~$L$ to be injected into~(\ref{eq_circular_curve}) to define a free circular trajectory at~$r$.

For a particle with mass~$m$, first denote~$\mathcal{E}$ and~$v$ the energy and spatial velocity measured by the ZAMO\footnote{\label{foot_E_ZAMO}For a massive particle,~$\mathcal{E} = \Gamma m$ where~$\Gamma$ is the Lorentz factor of the particle with respect to the ZAMO; for a massless particle,~$\mathcal{E} = h \nu$ where~$\nu$ is the frequency measured by the ZAMO.}, and~$n$ the 4-velocity of the latter, so that the 4-momentum of the particle decomposes as
\begin{eqnarray}
\label{eq_p_ZAMO}
p = \mathcal{E} (n + v) \text{ with } n \cdot v = 0.
\end{eqnarray}

One obtains
\begin{eqnarray}
\label{eq_E_L_E_ZAMO}
\mathcal{E} = \frac{E - \omega L}{N}
\end{eqnarray}
and
\begin{eqnarray}
\label{eq_spatial_velocity}
v = \frac{V}{Br} \partial_{\phi}
\end{eqnarray}
where
\begin{eqnarray}
\label{eq_signed_norm}
V = \frac{L}{Br\mathcal{E}}
\end{eqnarray}

Note that~$v^2 = V^2$, hence the name ``signed norm'' for~$V$.
Also recall here that a trajectory is defined to be prograde (resp. retrograde) when~$\omega L > 0$ (resp.~$\omega L < 0$); since~$\omega$ does not generally cancel, the~$\phi$ coordinate may be chosen such that~$\omega > 0$, which then simplifies in the definition.
Based on relations~(\ref{eq_spatial_velocity}) and~(\ref{eq_signed_norm}), and the fact that~$\omega(r_{\mathcal{H}})$ equals the angular velocity~$\Omega_{\mathcal{H}}$ of the event horizon%%%
\footnote{%%%
In quasi-isotropic coordinates, the event horizon is always located at a constant radial coordinate~$r_{\mathcal{H}}$.%%%
}%%%
, a prograde (resp. retrograde) trajectory intuitively rotates in the same (resp. opposite) direction as the black hole for the ZAMO.
However, for an observer at infinity, the angular velocity~$p^{\phi}/p^{t}$ of a circular orbit~(\ref{eq_circular_curve}) is
\begin{eqnarray}
\label{eq_angular_velocity}
\frac{p^{\phi}}{p^{t}} = \omega + \frac{NL}{Br\sqrt{(mBr)^{2} + L^{2}}},
\end{eqnarray}
so that one of the retrograde orbit may appear prograde from infinity if and only if it is outside the ergoregion and~$L > -m \omega (Br)^{2}/\sqrt{- N^{2} + (\omega Br)^{2}}$.

Note also that equation~(\ref{eq_circular_geod_condition_explicit}) (which is equivalent to~$\mathcal{V}' = 0$ as~$\mathcal{V} = 0$) is independent of~$m$ and hence homogeneous with respect to~$E$ and~$L$.
Therefore, injecting relations~(\ref{eq_E_L_E_ZAMO}) and~(\ref{eq_signed_norm}) into this expression allows to simplify all~$\mathcal{E}$, which finally yields the following second order equation in~$V$:
\begin{eqnarray}
\label{eq_V_circu}
\left( \frac{B'}{B} + \frac{1}{r} \right) V^{2} - \frac{B r \omega'}{N} V - \frac{N'}{N} = 0.
\end{eqnarray}

As announced, only the spatial velocity of the particle is constrained and not~$\mathcal{E}$, which means that, where a circular timelike geodesic exists, it can be the worldline of any massive particle regardless of its mass provided it has the right velocity, and where a circular null geodesic exists (i.e. a photon ring), it can be the worldline of any photon regardless of its frequency (this all seems consistent with the equivalence principle).
For a timelike circular geodesic to exist at~$r$, the values at~$r$ of the metric functions involved in~(\ref{eq_V_circu}) need to be such that at least one of the roots, if any exists, belongs to~$(-1,1)$ (the ZAMO must measure subluminal velocities).
In the massless case,~$V = \pm 1$ so that at any photon ring, if any exists, the metric functions need to be such that~$1$ or~$-1$ is a root of equation~(\ref{eq_V_circu}).
One then only needs to study the roots~$V_{\pm}$ of equation~(\ref{eq_V_circu}) to conclude about existence and location of timelike circular geodesics and photon rings.
These roots exist if and only if the discriminant
\begin{eqnarray}
\label{eq_D}
D = \left( \frac{B r \omega'}{N} \right)^{2} + \frac{4N'}{N} \left( \frac{B'}{B} + \frac{1}{r} \right)
\end{eqnarray}
of equation~(\ref{eq_V_circu}) is non-negative, in which case one has
\begin{eqnarray}
\label{eq_V_pm}
V_{\pm}(r) = \frac{\frac{B r \omega'}{N} \pm \sqrt{D}}{2\left( \frac{B'}{B} + \frac{1}{r} \right)}.
\end{eqnarray}

\begin{figure}
    \begin{center}
        \subfloat[Positivity of~$D$ allows (possibly superluminal) circular geodesics.]
        {
            \includegraphics[width=0.5\textwidth]{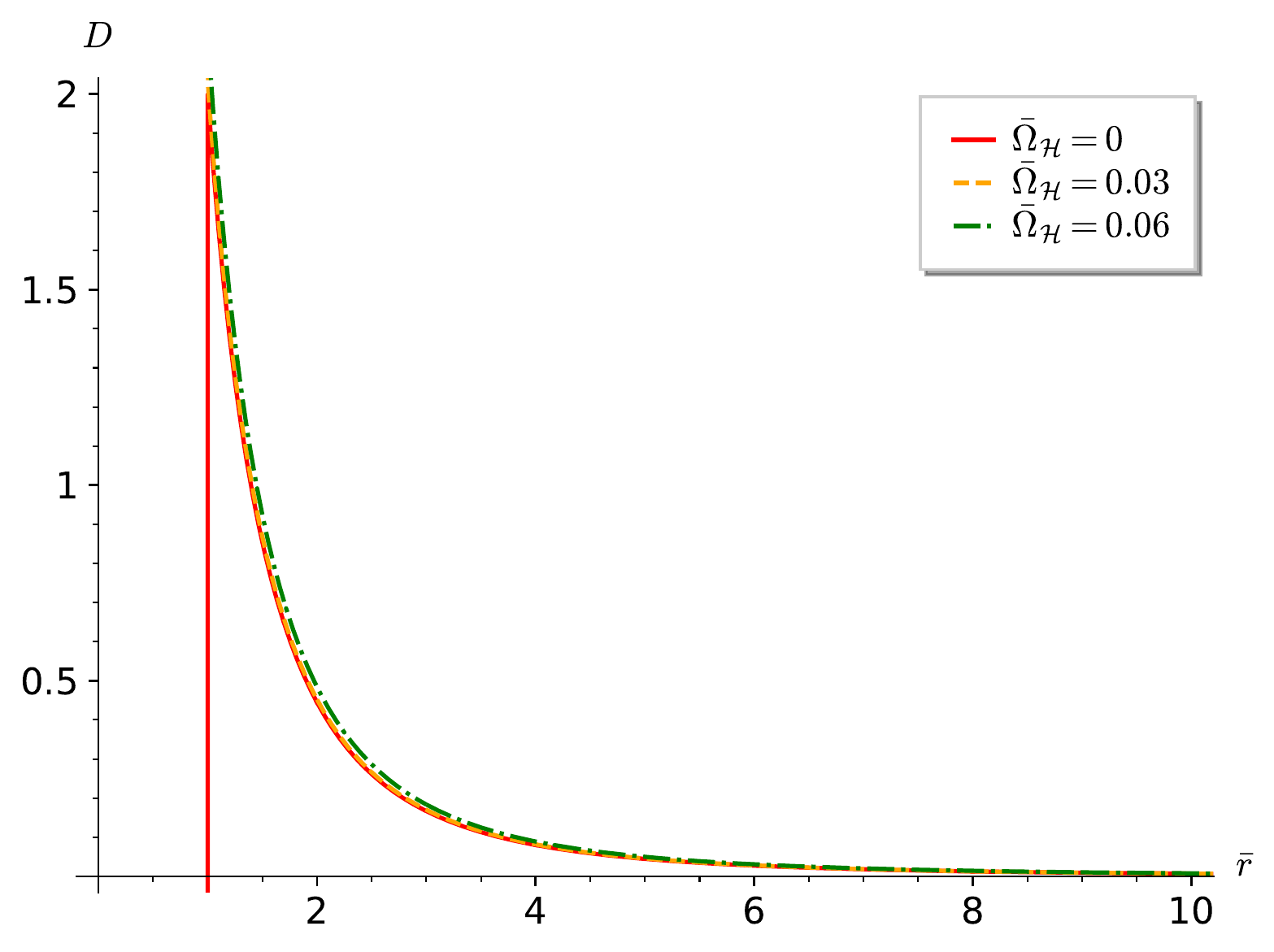}
            \label{fig_D_Kerr}
        }
        \subfloat[Velocities and photon rings.]
        {
            \includegraphics[width=0.5\textwidth]{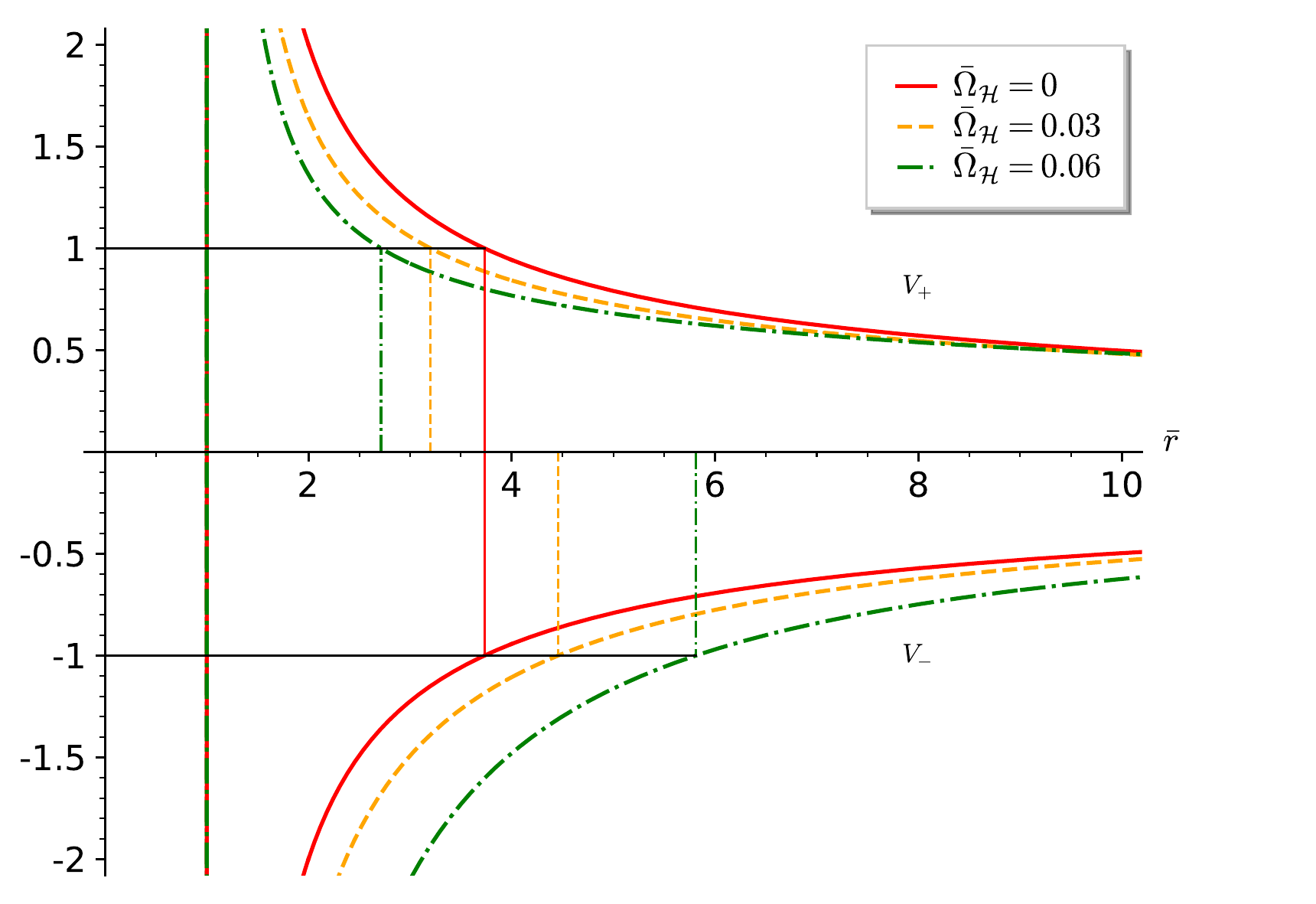}
            \label{fig_V_Kerr}
        } \\
        \subfloat[Lorentz factor of timelike circular geodesics.]
        {
            \includegraphics[width=0.5\textwidth]{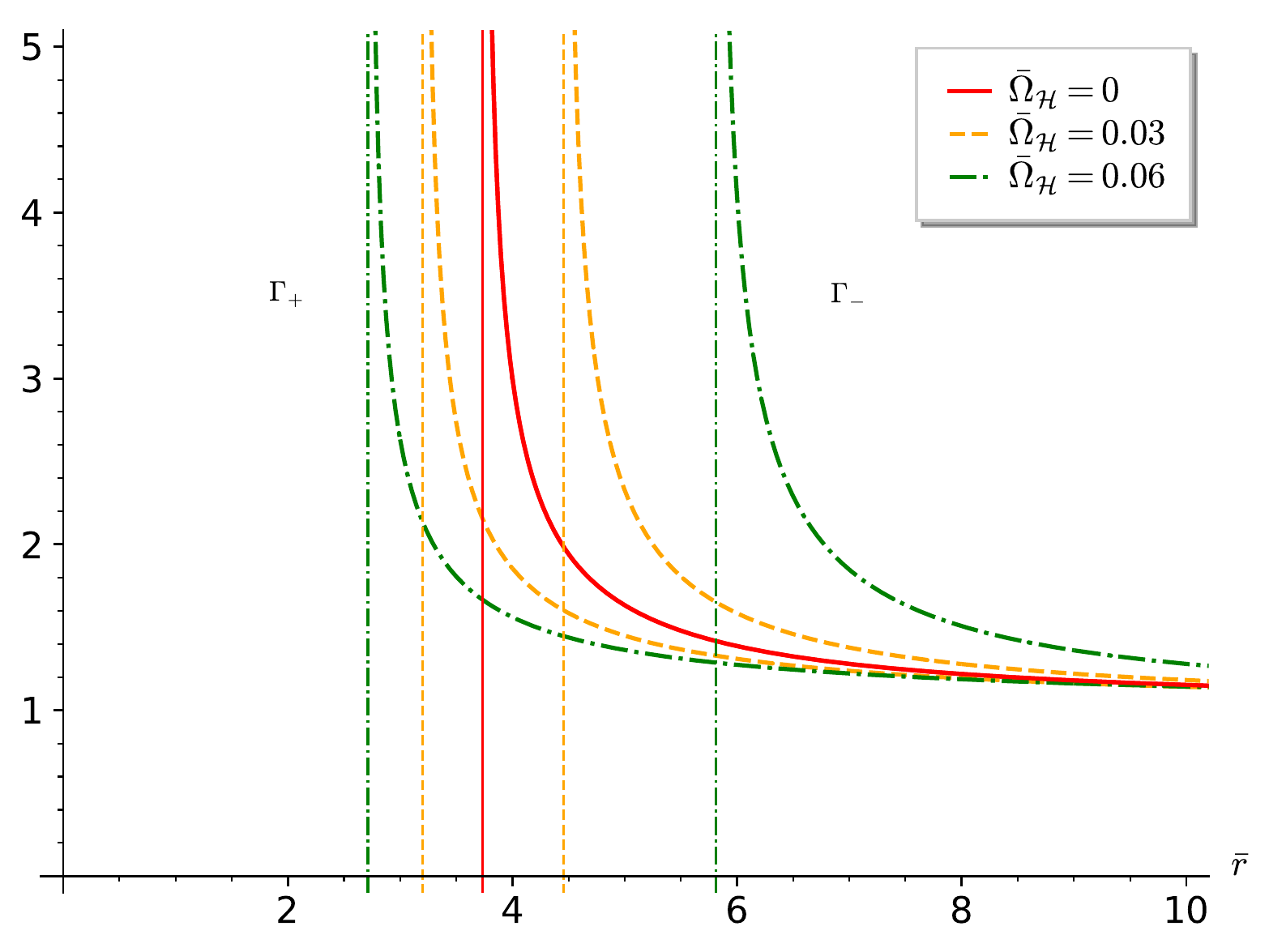}
            \label{fig_Gamma_Kerr}
        }
        \subfloat[Killing energy of timelike circular geodesics.]
        {
            \includegraphics[width=0.5\textwidth]{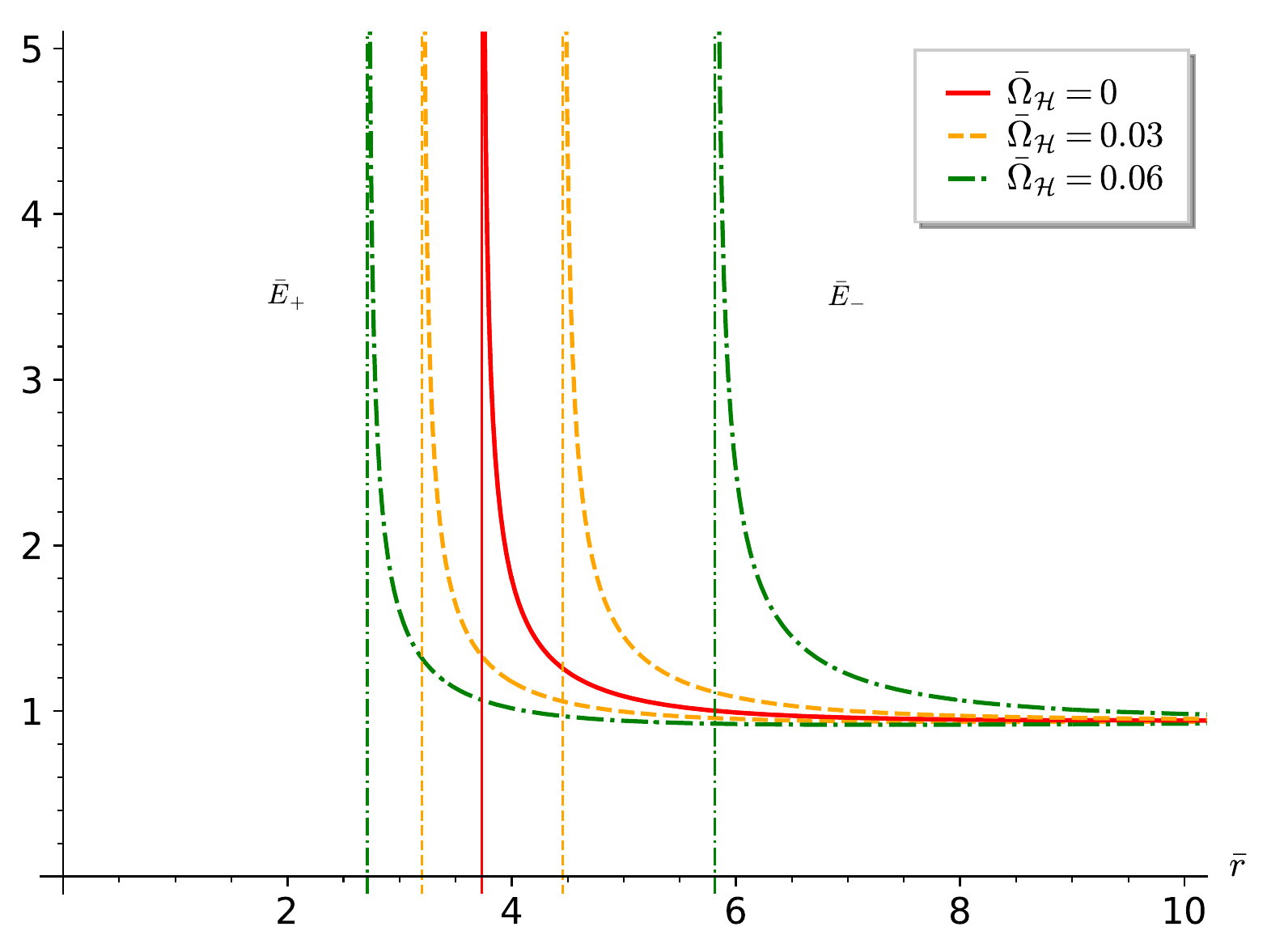}
            \label{fig_E_Kerr}
        } \\
        \subfloat[Killing angular momentum of timelike circular geodesics.]
        {
            \includegraphics[width=0.5\textwidth]{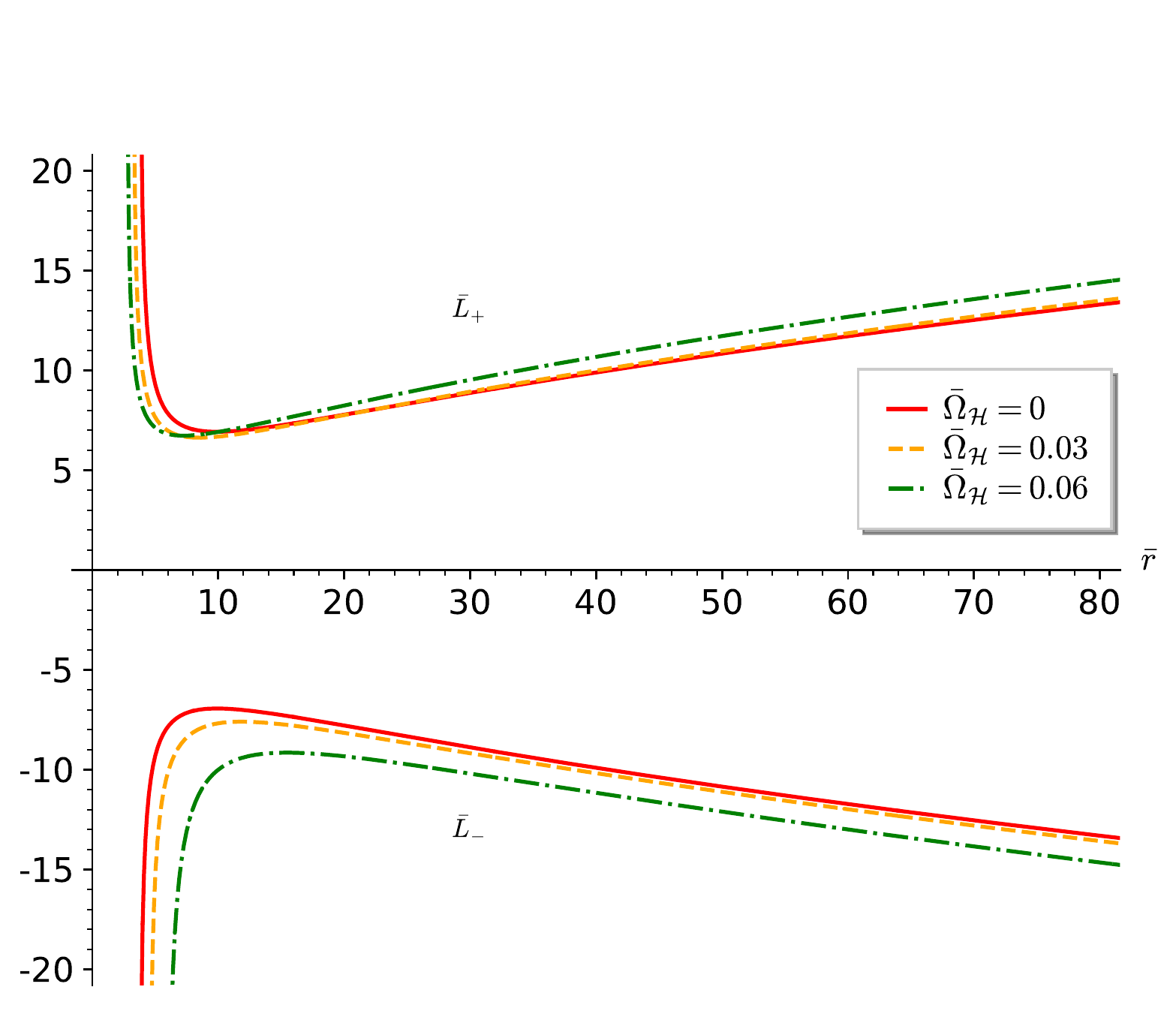}
            \label{fig_L_Kerr}
        }
    \end{center}
\caption{Kinematic characteristics of circular geodesics in Kerr spacetime for different dimensionless angular velocities~$\bar{\Omega}_{\mathcal{H}} = r_{\mathcal{H}} \Omega_{\mathcal{H}}$.}
\label{fig_Kerr}
\end{figure}

In Kerr spacetime, for any angular velocity~$\Omega_{\mathcal{H}}$%%%
\footnote{%%%
In Kerr spacetime, angular velocity~$\Omega_{\mathcal{H}}$ relates the usual mass~$M$ and spin parameter~$a$ according to~$\Omega_{\mathcal{H}} = a/(2M\left(M+\sqrt{M^{2}-a^{2}}\right))$.
}%%%
, function~$D$ monotonically decreases from
an infinite value at the horizon down to a zero limit as~$r \rightarrow +\infty$
(see figure~\ref{fig_D_Kerr}).
Therefore, each velocity function~$V_{\pm}$ necessarily becomes luminal at some point corresponding to a photon ring (marked with a vertical line from~$0$ to~$1$ in figure~\ref{fig_V_Kerr}), beyond which timelike circular geodesics exist everywhere.

Finally, relations~(\ref{eq_E_L_E_ZAMO}) and~(\ref{eq_signed_norm}) then yield
\begin{eqnarray}
\label{eq_L_circu_V}
L_{\pm} = \mathcal{E} B r V_{\pm}, \\
\label{eq_E_circu_V}
E_{\pm} = \mathcal{E} ( N + B r \omega V_{\pm} ),
\end{eqnarray}
i.e. in the massive case,
\begin{eqnarray}
\label{eq_L_massive_V}
L_{\pm} = \Gamma_{\pm} m B r V_{\pm}, \\
\label{eq_E_massive_V}
E_{\pm} = \Gamma_{\pm} m ( N + B r \omega V_{\pm} ),
\end{eqnarray}
with
\begin{eqnarray}
\label{eq_Lorentz}
\Gamma_{\pm} = (1 - V_{\pm}^{2})^{-1/2},
\end{eqnarray}
while in the massless case,
\begin{eqnarray}
\label{eq_L_massless_V}
L_{\pm} = \pm h \nu B r, \\
\label{eq_E_massless_V}
E_{\pm} = h \nu ( N \pm B r \omega ).
\end{eqnarray}

These quantities are plotted on figure~\ref{fig_Gamma_Kerr},~\ref{fig_E_Kerr} and~\ref{fig_L_Kerr} in the massive case for different dimensionless angular velocities~$\bar{\Omega}_{\mathcal{H}} = r_{\mathcal{H}} \Omega_{\mathcal{H}}$
in Kerr spacetime%%%
.
In particular, the ``+'' and ``-'' quantities are no longer merely equal or opposite in the rotating cases and respectively follow the evolution of the~``+'' and~``-''  photon rings in figure~\ref{fig_V_Kerr}: the~``+'' (resp.~``-'') Lorentz factor and Killing energy are for instance roughly shifted to the left (resp. right) of the Schwarzschild profile.

\section{Stability of circular geodesics}
\label{section_stability}

For circular geodesics, the radial equation~(\ref{eq_radial}) expectedly provides a stability criteria based on convexity.
A non-constant perturbation~$\delta$ (a constant perturbation would not threaten stability) to a circular geodesic at~$r$ allows to use equation~(\ref{eq_diff_eq_radial_noncircular}) instead:
\begin{eqnarray}
\label{eq_perturb}
\ddot{\delta} + \mathcal{V}'(r + \delta ,m,E + \delta_{E},L + \delta_{L}) = 0
\end{eqnarray}
in which Taylor expanding and invoking condition~(\ref{eq_circular_geod_condition}) yields
\begin{eqnarray}
\label{eq_perturb_o}
\ddot{\delta} + \mathcal{V}''(r,m,E,L)\delta = O(\delta_{E}) + O(\delta_{L}) + O(\delta^{2} + \delta_{E}^{2} + \delta_{L}^{2}).
\end{eqnarray}

If~$\mathcal{V}''(r,m,E,L) < 0$, then~$\delta$ must be accelerating away from~$r$ to preserve the asymptotic orders in the right-hand side.
To guarantee that any~$\delta$ is bounded in some neighbourhood of~$r$ then requires positive~$\mathcal{V}''(r,m,E,L)$.

Actually, the values of~$E$ and~$L$ for a circular geodesic at~$r$ are necessarily~$E_{\pm}(r)$ and~$L_{\pm}(r)$, explicitly given by relations~(\ref{eq_L_circu_V}) and~(\ref{eq_E_circu_V}).
In practice, one should thus study the sign of the two functions
\begin{eqnarray}
\label{eq_sign_stab}
\mathcal{V}_{\pm}'': r \mapsto \mathcal{V}''\left(r,m,E_{\pm}(r),L_{\pm}(r)\right),
\end{eqnarray}
on the set on which the discriminant~$D$ is non-negative.
Actually, the expressions~$\mathcal{V}''\left(r,m,E_{\pm}(r),L_{\pm}(r)\right)$ are homogeneous with respect to~$\mathcal{E}$, so that their sign do not depend on~$\Gamma_{\pm} m$ in the massive case nor on~$h \nu$ in the massless case.
Therefore, the stability of causal circular geodesics only depends on the sign of the two functions~(\ref{eq_sign_stab}) and concerns massive particles where~$V_{\pm}(r)~\in~(-1,1)$ and massless ones where~$V_{\pm}(r)~=~\pm 1$, regardless of whether the expressions used for~$E_{\pm}$ and~$L_{\pm}$ apply to a massive or a massless particle.

\setcounter{footnote}{0}

\begin{figure}
    \begin{center}
        \subfloat[Potentials~$\mathcal{V}(\cdot,m,E_{\pm}(r_{0}),L_{\pm}(r_{0}))$ of circular geodesics at various~$r_{0}$ in Schwarzschild spacetime.]
        {
            \includegraphics[width=0.5\textwidth]{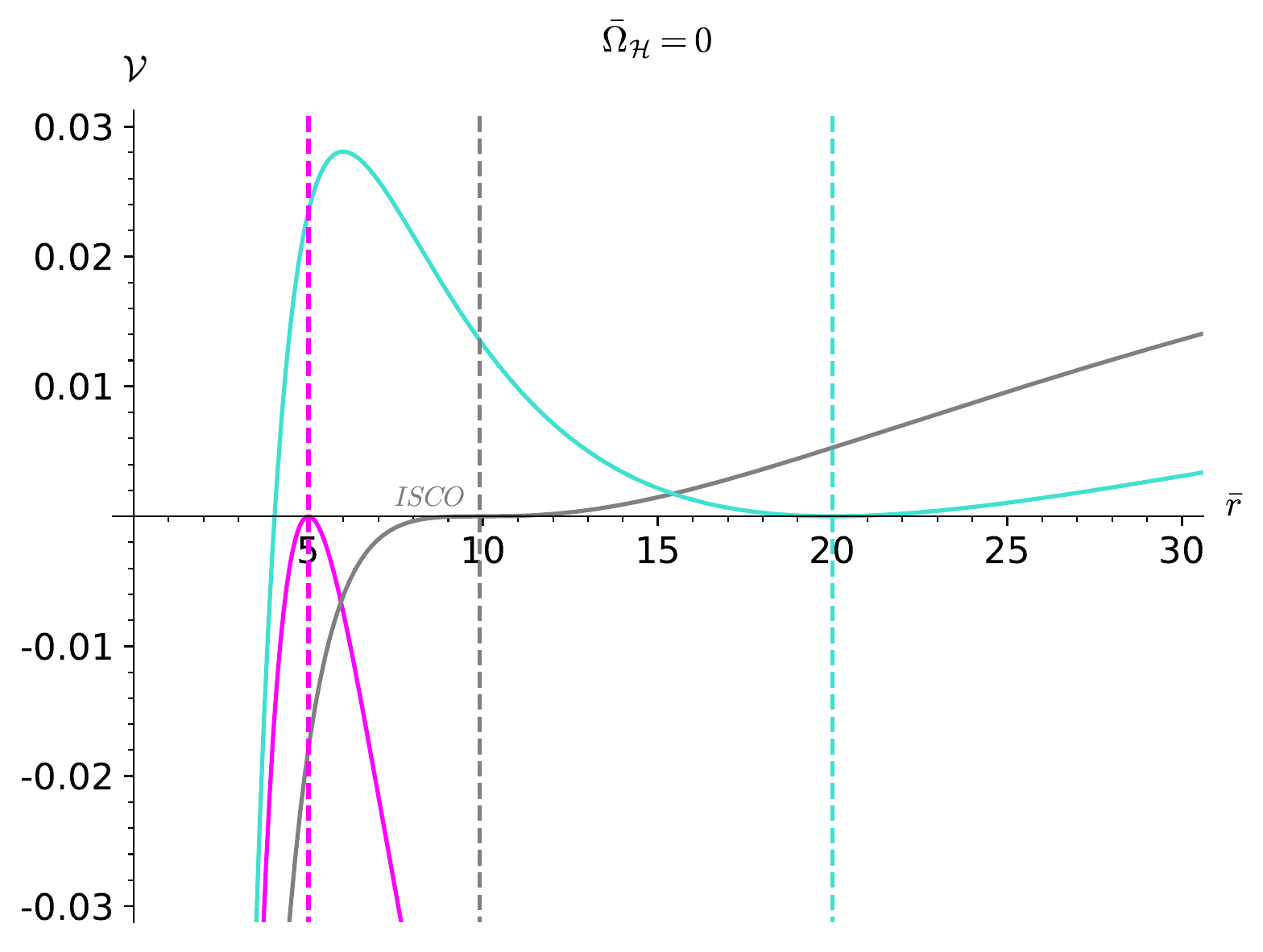}
            \label{fig_pots_circu_Schw}
        }
        \subfloat[$\mathcal{V}_{\pm}''(r_{0})$ is the convexity of~$\mathcal{V}(\cdot,m,E_{\pm}(r_{0}),L_{\pm}(r_{0}))$ evaluated at~$r_{0}$.]
        {
            \includegraphics[width=0.5\textwidth]{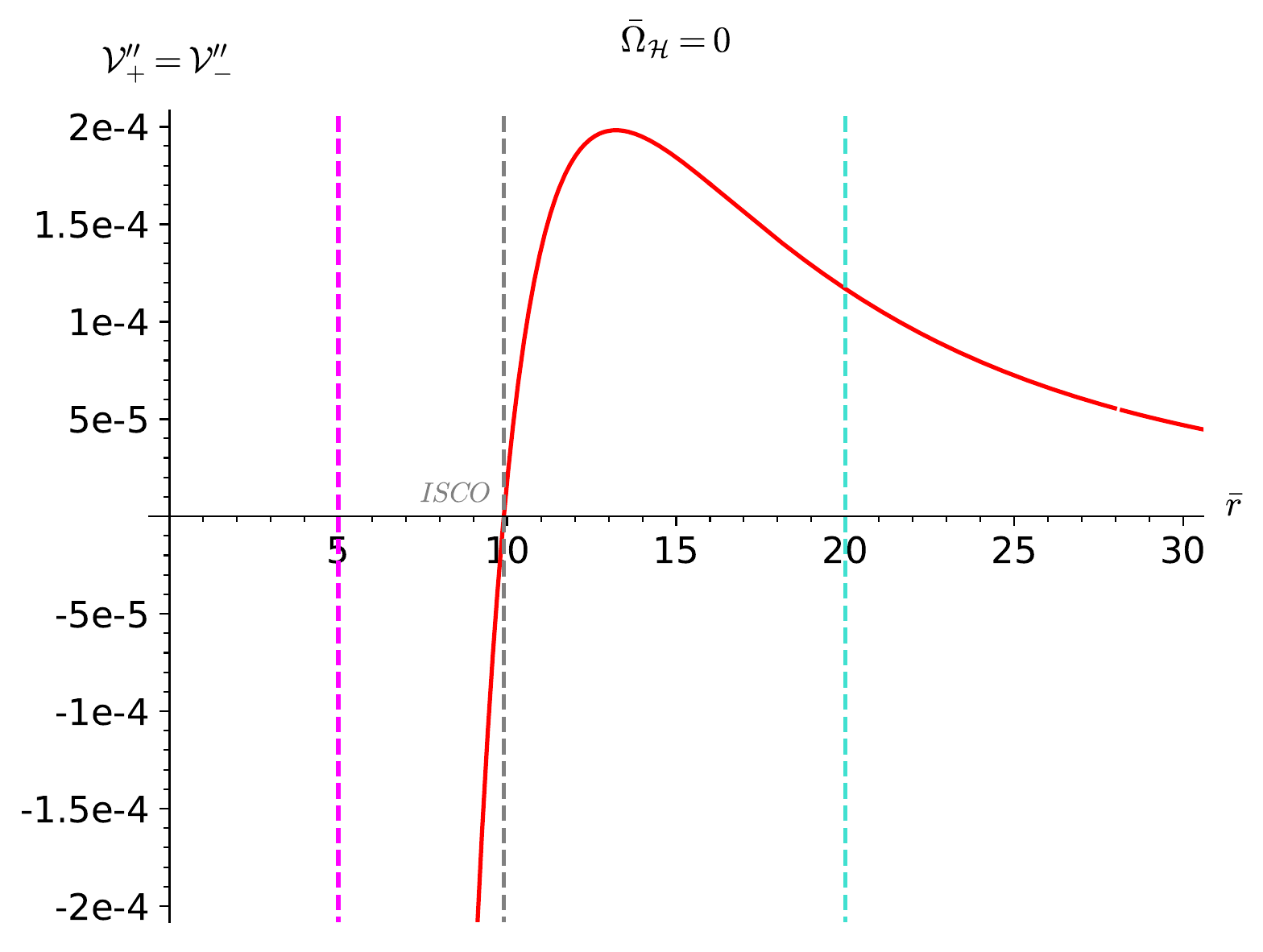}
            \label{fig_drrPot_Schw_ISCO}
        } \\
        \subfloat[Stability of circular geodesics from function~$\mathcal{V}_{\pm}''$.]
        {
            \includegraphics[width=0.5\textwidth]{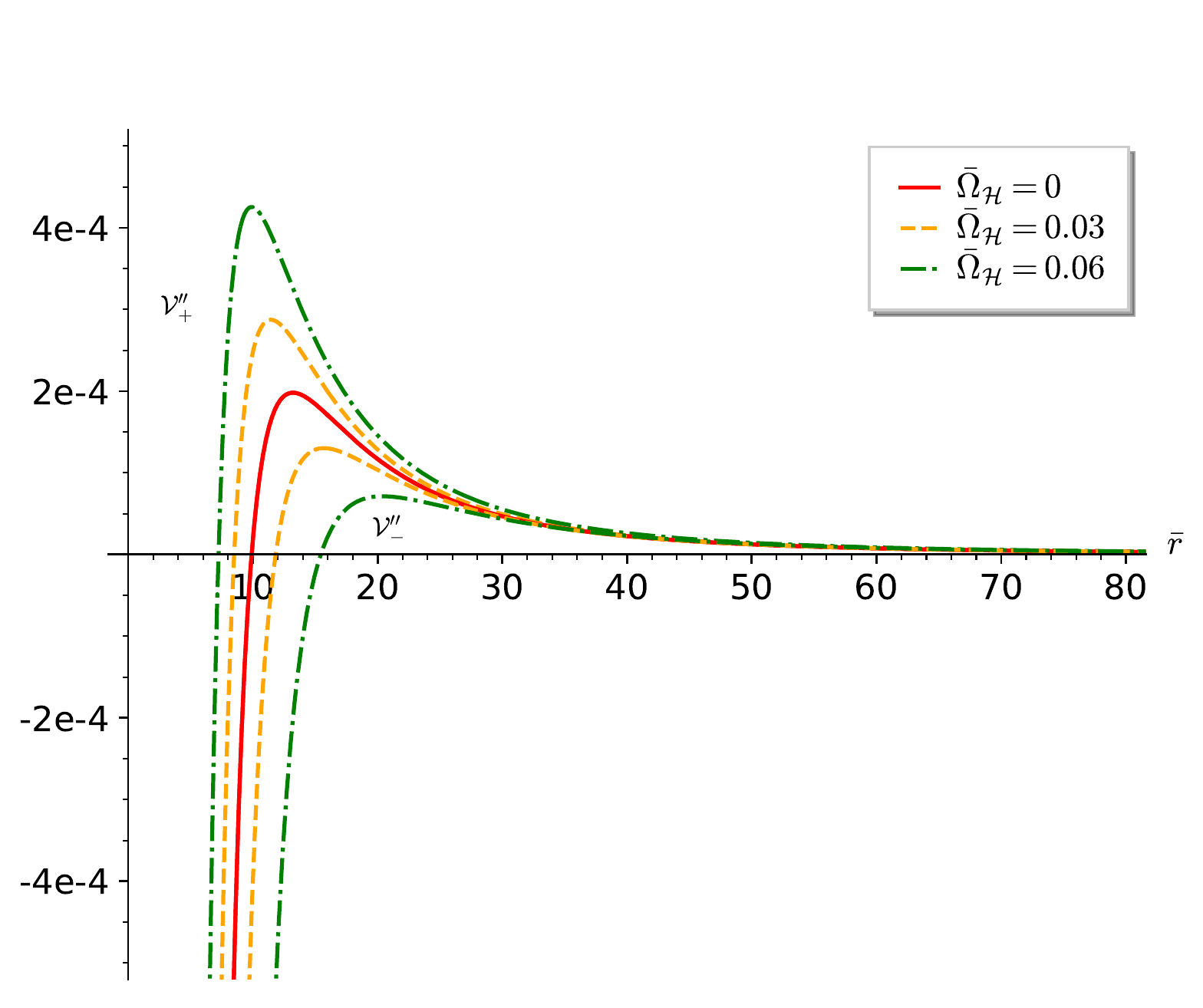}
            \label{fig_drrPot_Kerr}
        }
    \end{center}
    \caption{Positivity of~$\mathcal{V}_{\pm}''$ rules stability and hence location of ISCO.}
\label{fig_ISCO}
\end{figure}

Figures~\ref{fig_pots_circu_Schw} and~\ref{fig_drrPot_Schw_ISCO} illustrate all this in the Schwarzschild case: for each radii~$r_{0}$ marked with a dashed vertical line of a given colour in figure~\ref{fig_pots_circu_Schw}, the potential~$\mathcal{V}(\cdot,m,E_{\pm}(r_{0}),L_{\pm}(r_{0}))$, which corresponds to the geodesic at~$r_{0}$, is plotted with the same colour, and thus cancels in a stationary way at~$r_{0}$.
Then, for each~$r_{0}$, function~$\mathcal{V}_{\pm}''$ in figure~\ref{fig_drrPot_Schw_ISCO} extracts the convexity of~$\mathcal{V}(\cdot,m,E_{\pm}(r_{0}),L_{\pm}(r_{0}))$ at the corresponding~$r_{0}$: the purple curve is concave at~$r_{0} = 5r_{\mathcal{H}}$ on figure~\ref{fig_pots_circu_Schw} (hence unstable circular geodesic), so that~$\mathcal{V}_{\pm}''$ is negative at~$5r_{\mathcal{H}}$ on figure~\ref{fig_drrPot_Schw_ISCO}, whereas the turquoise curve is convex at~$r_{0} = 20r_{\mathcal{H}}$, so that~$\mathcal{V}_{\pm}''$ is positive at~$20r_{\mathcal{H}}$.
The limiting case (grey lines) such that~$\mathcal{V}(\cdot,m,E_{\pm}(r_{0}),L_{\pm}(r_{0}))$ cancels as an inflection point at~$r_{0} \simeq 9.9r_{\mathcal{H}}$ ($\mathcal{V}_{\pm}''(r_{0}) = 0$) defines the innermost stable circular orbit (ISCO)%%%
\footnote{%%%
In Boyer-Lindquist coordinates, the ISCO radius~$R_{0}$ of Schwarzschild spacetime is known to equal~$6M$; injecting this value into equation~(\ref{eq_r_QI}) does yield~$r_{0} = (5 + 2\sqrt{6}) M/2 \simeq 9.9 r_{\mathcal{H}}$.%%%
}%%%
.

Finally, figure~\ref{fig_drrPot_Kerr} gathers the functions~$\mathcal{V}_{\pm}''$ for different angular velocities.
The functions~$\mathcal{V}_{+}''$ and~$\mathcal{V}_{-}''$ are no longer equal in the rotating case~($\mathcal{V}_{+}''$ globally increases with rotation while~$\mathcal{V}_{-}''$ globally decreases) and thus respectively define an ISCO.
Based on figure~\ref{fig_V_Kerr}, both ISCO are always located beyond the corresponding photon ring, so that the latter are always unstable.

Note that for a stable circular geodesic~$(r_{0},m,E,L)$, the function~$\mathcal{V}(\cdot,m,E,L)$ realizes a local minimum (equal to zero) at~$r_{0}$, so that it is strictly positive in a neighbourhood of~$r_{0}$ except at~$r_{0}$.
Yet, one always has
\begin{eqnarray}
\label{eq_partial_E_Pot}
\partial_{E}\mathcal{V}(r_{0},m,E,L) = - \frac{\mathcal{E}}{N_{0} A_{0}^{2}} < 0
\end{eqnarray}
since~$\mathcal{E}$ is always strictly positive for a causal future-oriented curve.
Decreasing~$E$ thus increases~$\mathcal{V}$ locally\footnote{This is illustrated on figure~\ref{fig_pot_Kerr}:~$E$ is decreased to switch from the bottom curve to the top curve of a given color.}, so that~$\mathcal{V}$ becomes strictly positive on a neighbourhood of~$r_{0}$.
But~$\mathcal{V}$ is necessarily negative or zero on any orbit ruled by~(\ref{eq_radial}), which means that,~$m$ and~$L$ being fixed, there can be no geodesic close to~$(r_{0},m,E,L)$ with smaller~$E$: a stable circular geodesic at~$r_{0}$ realizes a local minimum of~$E$ on the set of geodesics having same~$m$ and~$L$.

\begin{figure}
    \begin{center}
    \includegraphics[width=0.8\textwidth]{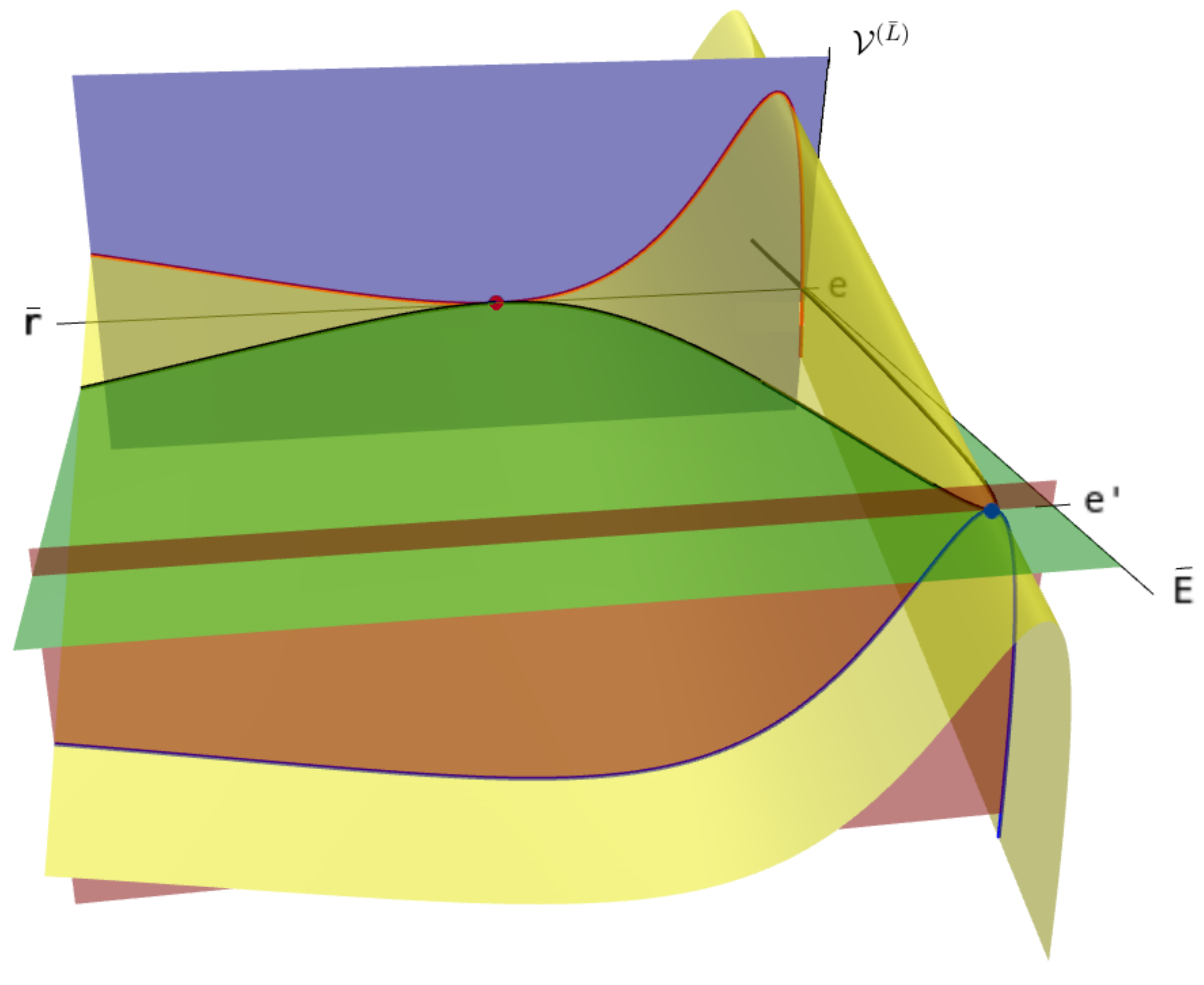}
    \end{center}
\caption{Effective potential~$\mathcal{V}^{(\bar{L})}$ (yellow surface) of massive particles at fixed~$\bar{L}$ in a Kerr spacetime~($a/M \simeq 0.52$). Its zeros lie on the black curve~$E^{(\bar{L})}_{min}$. Its profile at fixed Killing energy per unit mass~$e$ (resp.~$e'$) is highlighted as a red (resp. blue) curve on the blue (resp. red) vertical plane.~$\bar{L}$ being fixed, the only circular geodesics are marked with the red and blue dots, where both the two curves to which they respectively belong are stationary, and whose common convexity determines their stability.}
\label{fig_pot2vars_Kerr}
\end{figure}

As an additional note, fixing~$m$ and~$L$ also provides other interesting criteria to characterize circular geodesics among circular orbits (instead of~$\mathcal{V}'~=~0$), and investigate their stability (instead of~$\mathcal{V}'' > 0$).
For illustrative purposes,~$\mathcal{V}^{(m,L)}:~(r, E)~\mapsto~\mathcal{V}(r,m,E,L)$ is plotted for a Kerr spacetime in figure~\ref{fig_pot2vars_Kerr}.

From this point of view, circular geodesics lie on the set of points~$(r,E)$ such that~$\mathcal{V}^{(m,L)}(r,E) = 0$ (black horizontal curve in figure~\ref{fig_pot2vars_Kerr}): it is the intersection of the image of~$\mathcal{V}^{(m,L)}$ (yellow surface) with the~$\mathcal{V} = 0$ plane (green horizontal plane).
For any of these points~$(r_{0}, E_{0})$ to actually correspond to a circular geodesic (rather than an accelerated circular orbit), condition~(\ref{eq_circular_geod_condition}) must hold, meaning that~$r_{0}$ must be a stationary point of~$\mathcal{V}^{(m,L)}(\cdot,E_{0})$, whose graph is the intersection of the image of~$\mathcal{V}^{(m,L)}$ (yellow surface) with the~$E = E_{0}$ plane (blue or red vertical plane).

However, the set of points~$(r,E)$ such that~$\mathcal{V}^{(m,L)}(r,E) = 0$ (black horizontal curve) is always the graph of a function well-defined with respect to~$r$:~$r \mapsto E(r)$.
This comes from the fact that~$\mathcal{V}^{(m,L)}$ is a second order polynomial with respect to~$E$.
For any~$r$, the function~$\mathcal{V}^{(m,L)}(r,\cdot)$ always admits two distinct roots:
\begin{eqnarray}
\label{eq_E_min}
E^{(m,L)}_{min}(r) = \omega(r) L + N(r) \sqrt{m^{2} + \left(\frac{L}{B(r)r}\right)^{2}}, \\
\label{eq_E_neg}
E^{(m,L)}_{neg}(r) = - E^{(m,-L)}_{min}(r).
\end{eqnarray}

Since~$\mathcal{V}^{(m,L)}(r,\cdot)$ is a second order polynomial with negative dominant coefficient with respect to~$E$, and since~$E^{(m,L)}_{min}$ is always the greatest root, one has
\begin{eqnarray}
\label{eq_partial_E_Pot_E_min}
\partial_{E}\mathcal{V}^{(m,L)}\left(r,E^{(m,L)}_{min}(r)\right) < 0, \\
\label{eq_partial_E_Pot_E_neg}
\partial_{E}\mathcal{V}^{(m,L)}\left(r,E^{(m,L)}_{neg}(r)\right) > 0.
\end{eqnarray}

Based on equation~(\ref{eq_partial_E_Pot}),~$E^{(m,L)}_{neg}$ can in no case correspond to a causal future-oriented curve.
Actually, it merely corresponds to all the causal past-oriented circular and non-circular equatorial geodesics.
This had to be expected since the effective potential~$\mathcal{V}$ and the three other conservation equations can as well be used to describe them.
More precisely, based on relations~(\ref{eq_potential}),~(\ref{eq_radial_sqrt}),~(\ref{eq_t}) and~(\ref{eq_phi}), the past-oriented version of a future-oriented non-circular geodesic is obtained by switching the sign of both~$E$ and~$L$ and taking the opposite sign in~(\ref{eq_radial_sqrt}).
This is even easier to check in the circular case~(\ref{eq_circular_curve}), and relation~(\ref{eq_E_neg}) confirms that~$E^{(m,L)}_{neg}$ is merely the Killing energy of the past-oriented version of the circular orbit having opposite Killing angular momentum.
One may thus focus on~$E^{(m,L)}_{min}$ alone, which can only be negative for
retrograde orbits inside the ergoregion.

By definition, one has
\begin{eqnarray}
\label{eq_E_Pot_E_min}
\mathcal{V}^{(m,L)}\left(r,E^{(m,L)}_{min}(r)\right) = 0,
\end{eqnarray}
so that
\begin{eqnarray}
\label{eq_E_Pot_E_min_prime}
E^{(m,L)'}_{min}(r) = - \frac{\mathcal{V}^{(m,L)'}\left(r,E^{(m,L)}_{min}(r)\right)}{\partial_{E}\mathcal{V}^{(m,L)}\left(r,E^{(m,L)}_{min}(r)\right)}
\end{eqnarray}
where division is allowed by~(\ref{eq_partial_E_Pot_E_min}).
Therefore, the circular orbit~$(r,m,E^{(m,L)}_{min}(r),L)$ is a geodesic if and only if~$E^{(m,L)'}_{min}(r) = 0$ (instead of~$\mathcal{V}'(r,m,E^{(m,L)}_{min}(r),L) = 0$).
Graphically, one may check on figure~\ref{fig_pot2vars_Kerr} that at the circular geodesic marked with a red (resp. blue) dot, both the red curve~$\mathcal{V}^{(\bar{L})}(\cdot,e)$ (resp. blue curve~$\mathcal{V}^{(\bar{L})}(\cdot,e')$) and function~$E^{(\bar{L})}_{min}$ are stationary.

The second derivative of relation~(\ref{eq_E_Pot_E_min}) at a circular geodesic~$(r_{0},m,E^{(m,L)}_{min}(r_{0}),L)$ (to use~$E^{(m,L)'}_{min}(r_{0}) = 0$) merely yields
\begin{eqnarray}
\label{eq_E_Pot_E_min_prime_prime}
E^{(m,L)''}_{min}(r_{0}) = - \frac{\mathcal{V}^{(m,L)''}\left(r_{0},E^{(m,L)}_{min}(r_{0})\right)}{\partial_{E}\mathcal{V}^{(m,L)}\left(r_{0},E^{(m,L)}_{min}(r_{0})\right)}
\end{eqnarray}
which has the same sign as~$\mathcal{V}^{(m,L)''}(r_{0},E^{(m,L)}_{min}(r_{0}))$ because of~(\ref{eq_partial_E_Pot_E_min}).
Therefore, such circular geodesic is stable if and only if~$E^{(m,L)''}_{min}(r_{0}) > 0$ (instead of~$\mathcal{V}''(r_{0},m,E^{(m,L)}_{min}(r_{0}),L) > 0$).
Graphically, one may check on figure~\ref{fig_pot2vars_Kerr} that at the stable (resp. unstable) circular geodesic marked with a red (resp. blue) dot, both the red curve~$\mathcal{V}^{(\bar{L})}(\cdot,e)$ (resp. blue curve~$\mathcal{V}^{(\bar{L})}(\cdot,e')$) and function~$E^{(\bar{L})}_{min}$ are convex (resp. concave).

\begin{figure}
    \begin{center}
    \includegraphics[width=0.8\textwidth]{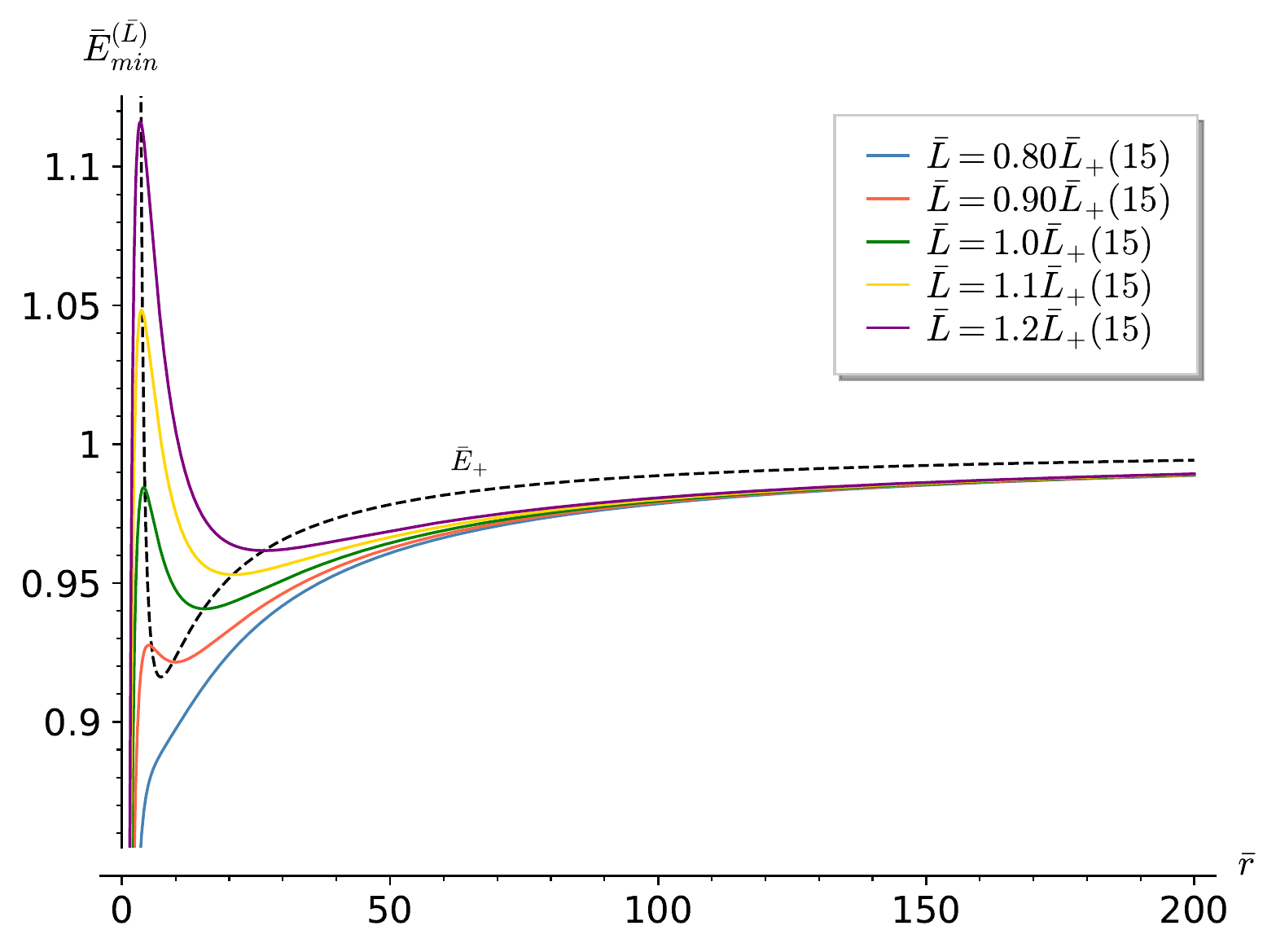}
    \end{center}
\caption{Function~$\bar{E}^{(\bar{L})}_{min}$ of massive particles for various positive~$\bar{L}$ in a Kerr spacetime~($a/M \simeq 0.52$).}
\label{fig_Emin_Kerr}
\end{figure}

It is thus possible to determine the location and stability of circular geodesics at fixed~$m$ and~$L$ from the function~$E^{(m,L)}_{min}$ rather than~$\mathcal{V}^{(m,L)}$.
In this case, one would then study the dependence of~$E^{(m,L)}_{min}$ on the parameter~$L$ (graphically, changing~$L$ deforms the yellow surface in figure~\ref{fig_pot2vars_Kerr} and hence the black curve).
Figure~\ref{fig_Emin_Kerr} illustrates this for a Kerr spacetime.
At fixed~$L = L_{0}$, circular geodesics correspond to the stationary points.
But then, for any stationary point~$(r_{0}, E_{0})$,~$E_{0}$ is necessarily equal to~$E_{\pm}(r_{0})$ (the sign depending on whether the orbit is prograde or retrograde) which is why the black dashed curve intersects each coloured curve at its stationary points.

Non-circular geodesics with Killing energy~$E$ are possible where~$E > E^{(m,L)}_{min}$.
Graphically, such a geodesic covers the region where the horizontal line~$E$ is above~$E^{(m,L)}_{min}$ while the abscissae of their intersections locate the periapsis and apoapsis of the geodesic.

Despite providing another interesting point of view on geodesics and stability criteria, using~$E^{(m,L)}_{min}$ to investigate stability is laborious since it requires to locate the circular geodesics and evaluate convexity for each~$L$.
Instead, it is much more efficient and exhaustive to focus on the sign of the two functions~$\mathcal{V}_{\pm}''$ given by~(\ref{eq_sign_stab}), as illustrated for Kerr spacetime with figure~\ref{fig_drrPot_Kerr} above.

For practical use, it is finally interesting to mention that the existence of a bounded non-circular orbit strongly suggests the existence of a stable circular orbit at some radius between its apsides, as one would expect intuitively.
Denote~$m$,~$E$ and~$L$ the mass, Killing energy and angular momentum of a particle following a bounded non-circular orbit such as the solid blue line in figure~\ref{fig_geods_Kerr} ruled by the right, negative, bounded branch of the bottom blue curve of figure~\ref{fig_pot_Kerr}.
The argument is that~$\mathcal{V}(\cdot,m,E,L))$ necessarily admits a local minimum at some~$r_{0}$ between its apsides, i.e. such that
\begin{eqnarray}
\label{eq_V_0}
\mathcal{V}(r_{0},m,E,L) < 0, \\
\label{eq_Vp_0}
\mathcal{V}'(r_{0},m,E,L) = 0, \\
\label{eq_Vpp_0}
\mathcal{V}''(r_{0},m,E,L) > 0.
\end{eqnarray}

Since~$\mathcal{V}(r,m,0,0) > 0$ and~$\mathcal{V}(r,m,E,L) < 0$, there necessarily exists a strictly positive factor~$\alpha$ rescaling~$E$ and~$L$ in such a way that
\begin{eqnarray}
\label{eq_pot_alpha_0}
\mathcal{V}(r,m,\alpha E,\alpha L) = 0.
\end{eqnarray}
Defining
\begin{eqnarray}
\label{eq_X}
\mathcal{X}(r,m,E,L) = \frac{1}{2A^{2}} \left[ - \left( \frac{E - \omega L}{N} \right)^{2} + \left( \frac{L}{B r} \right)^{2} \right]
\end{eqnarray}
so that~$\mathcal{V} = 1/(2A^{2}) + \mathcal{X}$, one has, for any~$\beta$,
\begin{eqnarray}
\label{eq_Xp}
\mathcal{V}'(r_{0},m,\beta E,\beta L) \approx \mathcal{X}'(r_{0},m,\beta E,\beta L) = \beta^{2} \mathcal{X}'(r_{0},m,E,L), \\
\label{eq_Xpp}
\mathcal{V}''(r_{0},m,\beta E,\beta L) \approx \mathcal{X}''(r_{0},m,\beta E,\beta L) = \beta^{2} \mathcal{X}''(r_{0},m,E,L)
\end{eqnarray}
far from the strong-field region (so that~$A$ slowly varies) where most stars orbiting a central object would be observed.
The quantities~(\ref{eq_Xp}) and~(\ref{eq_Xpp}) cancel for~$\beta~=~1$ according to~(\ref{eq_Vp_0}) and~(\ref{eq_Vpp_0}), which implies the same for~$\beta=\alpha$.
The parameters~$(r_{0},m,\alpha E,\alpha L)$ are thus close to define a stable circular orbit, and it should generally be possible to finish tuning them to obtain an exact stable circular orbit.
This makes the previously discussed stability criteria for circular orbits relevant for observations although no exactly circular orbit exists; this means for instance that observing a star on a non-circular orbit around a black hole logically requires the ISCO to be located below the apoapsis of the star orbit.

\section*{Acknowledgements}

The author thanks Eric Gourgoulhon for helpful discussions and comments on this paper.
This work was supported by the CNRS project 80PRIME-TNENGRAV.

\appendix

\section{Kerr metric in quasi-isotropic coordinates}
\label{appdx_Kerr_QI}

Denoting~$M$ the mass and~$a$ the spin parameter, the four metric functions involved in the quasi-isotropic expression~(\ref{eq_circular_spacetime}) of the Kerr metric explicitly write as
\begin{eqnarray}
N^{2} = \frac{\Sigma\Delta}{\Sigma(R^{2}+a^{2}) + 2a^{2}MR\sin^{2}\theta},    \\
A^{2} = \frac{\Sigma}{r^{2}},    \\
B^{2} = \frac{1}{r^{2}}\left( R^{2}+a^{2} + \frac{2a^{2}MR\sin^{2}\theta}{\Sigma}\right), \\
\omega = \frac{2aMR}{\Sigma(R^{2}+a^{2}) + 2a^{2}MR\sin^{2}\theta},
\end{eqnarray}
where
\begin{eqnarray}
\label{eq_R_BL}
R = r + \frac{M^{2} - a^{2}}{4r} + M,   \\
\Sigma = R^{2} + a^{2} \cos^{2}\theta,  \\
\Delta = R^{2} + a^{2} - 2MR.
\end{eqnarray}

The function~$R$ defined by relation~(\ref{eq_R_BL}) actually corresponds to the radial coordinate of the Boyer-Lindquist system (which is presented e.g. in section~33.2
of~\cite{MTW}).
It is inverted as
\begin{eqnarray}
\label{eq_r_QI}
r = \frac{1}{2} \left( R + \sqrt{R^{2} - 2MR + a^{2}} - M \right).
\end{eqnarray}

The remaining coordinates are identical in the quasi-isotropic and Boyer-Lindquist systems.

\section*{References}

\bibliographystyle{iopart-num}
\bibliography{biblio}

\end{document}